\documentclass[submission,copyright,creativecommons,sort&compress]{eptcs}
\providecommand{\event}{FROM 2025} 

\usepackage{mathtools}
\usepackage{amsthm}
\usepackage{amssymb, amsmath}
\usepackage{multirow}
\usepackage{array}
\usepackage{enumitem}
\usepackage[square, numbers]{natbib}
\usepackage{cleveref}
\usepackage{ebproof}
\usepackage{subcaption}
\usepackage{xparse}
\usepackage{etoolbox}

\crefformat{footnote}{#2\footnotemark[#1]#3}

\usepackage{iftex}

\MakeRobust{\ref}

\makeatletter
\newcommand{\labeltext}[2]{%
  #1
  \@bsphack
  \csname phantomsection\endcsname 
  \def\@currentlabel{#1}{\label{#2}}%
  \@esphack
}
\makeatother

\ifpdf
  \usepackage{underscore}         
  \usepackage[T1]{fontenc}        
\else
  \usepackage{breakurl}           
\fi

\def\titletext{On a Dependently Typed Encoding of Matching Logic}

\title{\titletext}
\author{Ádám Kurucz
\institute{ELTE Eötvös Loránd University\\
Budapest, Hungary}
\email{cphfw1@inf.elte.hu}
\and
Péter Bereczky
\institute{ELTE Eötvös Loránd University\\
Budapest, Hungary}
\email{berpeti@inf.elte.hu}
\and
Dániel Horpácsi
\institute{ELTE Eötvös Loránd University\\
Budapest, Hungary}
\email{daniel-h@elte.hu}
}

\DeclareUnicodeCharacter{03BC}{\ensuremath{\mu}}

\newcommand{\refl}{\textit{refl}}
\newcommand{\Sorts}{\textit{Sorts}}
\newcommand{\Type}{\textit{Type}}
\newcommand{\ld}{.\ }
\newcommand{\lfp}{\textbf{lfp}}
\newcommand{\nat}{\mathbb N}
\newcommand{\bool}{\mathbb B}
\newcommand{\Coq}{Coq}

\newcommand{\argsorts}{params}
\newcommand{\retsorts}{return}
\newcommand{\placeholder}{\square}
\newcommand{\emptyhetlist}{\langle\rangle}

\NewDocumentCommand{\sepend}{m m m s}{#3\IfBooleanTF{#4}{#2}{#1\sepend{#1}{#2}}}
\newcommand{\sep}[1]{\sepend{#1}{}}
\newcommand{\polyapp}[1]{#1(\sepend{,\ }{)}}
\newcommand{\app}{\sep{\ }}
\newcommand{\sepcomma}{\sep{,\ }}
\newcommand{\seplist}{[\sepend{,\ }{]}}
\newcommand{\sephetlist}{\langle\sepend{,\ }{\rangle}}

\newcommand{\FV}[1]{FV(#1)}
\newcommand{\ofsortvar}[2]{{#1}:{#2}}
\newcommand{\ofsort}[2]{#1_{#2}}
\newcommand{\update}[3]{#1[#2 \mapsto #3]}

\newcommand{\subst}[3]{#1[#3/#2]}
\NewDocumentCommand{\defined}{m o O{#2}}{\lceil #1 \rceil \IfValueT{#2}{_{#2}^{#3}}}
\NewDocumentCommand{\total}{m o O{#2}}{\lfloor #1 \rfloor \IfValueT{#2}{_{#2}^{#3}}}
\newcommand{\bevar}[1]{\underline{#1}}
\newcommand{\bsvar}[1]{\underline{\underline{#1}}}
\newcommand{\fevar}[1]{\widehat{#1}}
\newcommand{\fsvar}[1]{\widehat{\widehat{#1}}}
\newcommand{\existentialsort}[2]{#1_{#2}}

\newcommand{\syapp}[2]{#1 \cdot #2}

\newcommand{\sortedpattern}[1]{\app {\texttt{Pattern}} {#1} *}
\newcommand{\closedpattern}[2]{\app {\texttt{Pattern}} {#1} {#2} *}
\newcommand{\sortedclosedpattern}[3]{\app {\texttt{Pattern}} {#2} {#3} {#1} *}

\newcommand{\doubleplus}{\mathbin{{+}\mspace{-5mu}{+}}}
\newcommand{\concat}{\doubleplus}
\newcommand{\fmap}{\mathbin{<\mspace{-6mu}\$\mspace{-6mu}>}}
\newcommand{\map}[2]{#1 \fmap #2}

\newcommand{\listof}[1]{\app {\textit{List}} {#1} *}

\NewDocumentCommand{\EVar}{o}{\IfValueTF{#1}{\ofsort{EV}{#1}}{EV}}
\NewDocumentCommand{\SVar}{o}{\IfValueTF{#1}{\ofsort{SV}{#1}}{SV}}
\newcommand{\nth}[1][n]{\ensuremath{#1^\text{th}}}

\NewCommandCopy{\foexists}{\exists}
\RenewDocumentCommand{\exists}{o}{\IfValueTF{#1}{\existentialsort{\foexists}{#1}}{\foexists}}

\theoremstyle{definition}
\newtheorem{definition}{Definition}

\makeatletter

\newcommand{\superimpose}[2]{{%
  \ooalign{%
    \hfil$\m@th#1\@firstoftwo#2$\hfil\cr
    \hfil$\m@th#1\@secondoftwo#2$\hfil\cr
  }%
}}

\NewDocumentCommand{\typedef@@line}{m m m m}{
	#2 & {} : #3 \\
	\IfValueT{#4}{& \hfill #4 \\}
}
\NewDocumentCommand{\typedef@line}{t? >{\SplitArgument{2}{!}}m s}{
	\typedef@@line {#1} #2
	\IfBooleanF{#3}{\typedef@line}
}
\NewDocumentEnvironment{typedef}{m s o >{\SplitList{;}}+b}{
	\IfBooleanTF{#2}{
		\def\typename{\textit{#1}}
	}{
		\def\typename{\texttt{#1}}
	}
	\center
	\tabular{>{\(}l<{\)} @{\hspace{0.5em}} >{\(}l<{\)}}
	\IfValueT{#3}{\typename & {} : #3 \\}
	\typedef@line #4 *
}{
	\endtabular
	\endcenter
}
\makeatother

\makeatletter
\newcommand{\pushright}[1]{\ifmeasuring@#1\else\omit\hfill$\displaystyle#1$\fi\ignorespaces}
\makeatother

\def\titlerunning{\titletext}
\def\authorrunning{Á. Kurucz, P. Bereczky, D. Horpácsi}

\begin{document}

\maketitle

\begin{abstract}
Matching logic is a general formal framework for reasoning about a wide range of theories, with particular emphasis on programming language semantics. Notably, the intermediate language of the $\mathbb{K}$ semantics framework is an extension of matching $\mu$-logic, a sorted, polyadic variant of the logic. Metatheoretic reasoning requires the logic to be expressed within a foundational theory; opting for a dependently typed one enables well-sortedness in the object theory to correspond directly to well-typedness in the host theory. In this paper, we present the first dependently typed definition of matching $\mu$-logic, ensuring well-sortedness via sorted contexts encoded in type indices. As a result, ill-sorted syntax elements are unrepresentable, and the semantics of well-sorted elements are guaranteed to lie within the domain of their associated sort.
\end{abstract}

\section{Introduction}

Matching logic is a formal framework that generalizes first-order logic. The fundamental building block of the logic is the \emph{pattern}, which serves the same purpose as both predicates and terms in first-order logic.
Patterns are interpreted over a predefined carrier set, and their semantics is the set of elements that ``match'' them. The various pattern constructors extend their meanings from first-order logic; for example, negation is associated with the set complement operation, while conjunctions are interpreted as the intersection of the corresponding sets of the parameters.

Various theories have been axiomatized in matching logic. Basic examples include standard types such as booleans, natural numbers, or lists, as well as the theory of \emph{definedness}~\cite{chen2021explained}, which introduces the concept of equality. Furthermore, temporal logic can be formalized within this framework, thereby enabling the modeling of rewrite systems. One system that utilizes a rewriting logic is the $\mathbb{K}$ framework, which can be used to describe programming language syntax and semantics.
Notably, all $\mathbb{K}$ definitions can be translated into matching logic theories, facilitating reasoning about program correctness~\cite{chen2019mu}.

However, the consistency of these theories is not verified. Based on inconsistent theories any statement 
may be proven, compromising the trustworthiness of the system.
Consistency can be checked via the soundness theorem of the logic: if a model can be given for a theory then it is necessarily consistent.
We aim to establish a framework in which the consistency of matching logic theories can be verified this way, using metatheoretic reasoning.
To this end, we formalize the logic in a way that permits the description of theories, models for them, and the verification that these models satisfy the theories.

While several formalizations exist for matching logic~\cite{ml-in-mm,aml-in-coq-traian,aml-in-coq-unif,aml-harp,aml-lean}, those express unsorted, applicative variants of the logic (defining sortedness as a theory), whereas matching $\mu$-logic, the foundation behind the $\mathbb{K}$ framework is sorted and polyadic (i.e., it features $n$-ary applications instead of binary ones).\footnote{The $\mathbb{K}$ framework's intermediate language, Kore, is an extension of matching $\mu$-logic.}
These properties, especially sort checking, bring additional complexity in the formalization, but most of it can be outsourced to the metatheory, while the lack of partial applications and ill-sorted patterns facilities simpler reasoning about models and consistency.

In this paper, we present an encoding of matching $\mu$-logic within a dependent type theory featuring inductive definitions. A key contribution of our work lies in the shallow embedding of sorting and variable contexts. In this approach, the typing rules of the metatheory ensure the well-formedness of patterns, eliminating the need of explicit reasoning about well-sortedness and closedness. Of particular interest is the encoding of well-sorted substitution, which necessitates both the adaptation of an existing substitution calculus and the embedding of this calculus within the metatheory.
Furthermore, unlike in related work, we include \emph{definedness} in the logic, which makes frequently used logical constructs (such as equality and set inclusion) expressable using primitives, ultimately simplifying semantics-based reasoning about concrete patterns.
As an implementation of the dependently typed metatheory, we have chosen to use the \Coq{} proof assistant~\cite{coq}\footnote{Starting at version 9.0, the proof assistant was renamed to Rocq, however, this implementation uses version 8.20, which is still officially called Coq. Because of that, we use this name to refer to the system.}. The machine-checked formalization is available at~\cite{kore-ml}.

The rest of this paper is structured as follows. \Cref{sec-background} introduces various metatheoretical concepts along with matching logic. Next, \Cref{sec-work} summarizes related work. Thereafter, \Cref{sec-syntax} introduces the dependently typed syntax of matching logic, including substitutions, and \Cref{sec-semantics} discusses the dependently typed semantics. \Cref{sec-conclusion} discusses future directions and concludes.

\section{Background}\label{sec-background}

This section provides the necessary background for this paper, including a brief introduction to dependent type theory and matching logic.

\subsection{Common metatheoretical types}\label{subsec-metatypes}

First, we show some well-known inductive types in the metatheory, demonstrating the notations we will use throughout the paper. Note that in the metatheory, we use curried syntax for function application.

\paragraph{Natural numbers and lists} To represent natural numbers, we use Peano's axiomatization, with constructors called $O$ (denoting zero) and $S$ (denoting the successor). For lists, we use algebraic linked lists, where $nil$ represents the empty list and $cons$ is for prepending values to a list.

\medskip

\noindent\begin{minipage}{0.4\textwidth}\centering
\begin{typedef}{$\mathbb N$}[\Type]
	O ! \typename;
	S ! \typename \to \typename
\end{typedef}
\end{minipage}%
\begin{minipage}{0.60\textwidth}\centering
\begin{typedef}{List}*[\Type \to \Type]
	\textit{nil} ! \forall (A : \Type) \ld \app {\typename} A *;
	\textit{cons} ! \forall (A : \Type) \ld A \to \app {\typename} A * \to \app {\typename} A *
\end{typedef}
\end{minipage}

\medskip

We will omit the \Type{} argument of the constructors whenever they can be inferred from the context. We define addition on natural numbers, denoted $a + b$ in the usual manner, with recursive descent on the left argument.  For $cons$, we use the right associative infix symbol $::$, where $\sep {::} a b {nil} *$ means $\app {cons} a {(\app {cons} b {nil} *)} *$. Furthermore, we use $\seplist a b c *$ to mean $\sep{::} a b c {nil} *$, and denote the concatenation of $xs$ and $ys$ by $xs \concat ys$. 
Finally, for some types $A$ and $B$, function $f : A \to B$ and list $xs : \listof{A}$, the expression $\map f {xs}$ denotes a value of type $\listof{B}$ which we obtain by applying the function $f$ to every element of $xs$.

\subsection{Dependent type theory}\label{subsec-dependenttt}

In dependent type theory, types can depend on concrete values of other types~\cite{chipala2022certified}.
As shown previously, in a non-dependent setting, we may define types representing simple data such as natural numbers or lists. Note however, that while the latter may be parametric in other types, we have no way of describing lists of a particular length, for example, as that would depend on a natural number. This is made possible by dependent types.

\paragraph{Vectors} In fact, the type of lists with a given length (also known as vectors) is commonly used in dependently typed languages and proof assistants. Similar to lists, it has two constructors, one that represents empty vectors and has a length of $0$, and one that prepends an element to the vector, incrementing its length:

\begin{typedef}{Vec}[\Type \to \mathbb N \to \Type]
	\textit{vnil} ! \forall (A : \Type) \ld \app {\typename} A 0 *;
	\textit{vcons} ! \forall (A : \Type)\ (n : \mathbb N) \ld A \to \app {\typename} A n * \to \app {\typename} A {(\app S n *)} *
\end{typedef}

\noindent
The type of \texttt{Vec} is $\Type \to \mathbb N \to \Type$, which clearly shows that, besides a type, it also depends on a value of type $\mathbb N$. These type parameters are often referred to as (type) indices. Going forward, we will say that vectors ``are indexed by'' types and natural numbers to refer to this signature.

\paragraph{Heterogeneous lists} The \texttt{Vec} type may be used in place of $n$-ary tuples, however, it is not enough for certain purposes, such as grouping an $n$-ary function's arguments into a list of length $n$ (called uncurrying) because these arguments may have different types. For this, we define a generalization of vectors, called heterogeneous lists, which are indexed by a list of $\Type$s that they must contain, in order.
A $\app {\texttt{Vec}} A n *$ type may be expressed as a \texttt{HList}, with its index list containing the $A$ type $n$ times.

\begin{typedef}{HList}[\listof{\Type} \to \Type]
    \textit{hnil} ! \typename\ [];
	\textit{hcons} ! \forall (A : \Type)\ (\textit{As} : \listof{\Type}) \ld A \to \app {\typename} {\textit{As}} * \to \app {\typename} {(A :: \textit{As})} *
\end{typedef}

\noindent
We use the syntactic sugar $\sephetlist a b c *$ for heterogeneous lists, to mean $\app {\textit{hcons}} a {(\app {\textit{hcons}} b {(\app {\textit{hcons}} c {\textit{hnil}} *)} *)} *$.

\paragraph{Dependent pair} Another commonly used dependent type is the dependent pair. This is like the regular pair, except that the type of the second element depends on the value of the first element:

\begin{typedef}{DPair}[\forall(A : \Type) \ld (A \to \Type) \to \Type]
	\textit{dpair} ! \forall (A : \Type)\ (P : A \to \Type)\ (x : A) \ld \app P x * \to \app {\typename} A P *
\end{typedef}

\noindent
We reuse the notation of regular pairs for dependent ones as well, writing $(x,\ y)$ in place of $\app {\textit{dpair}} x y *$.
As an example usage of these pairs, consider the type $\app {\texttt{DPair}} {(\listof{\Type})} {\texttt{HList}} *$, which represents all heterogeneous lists. Valid values of this type include $([],\ \emptyhetlist)$ and $(\seplist {\mathbb N} {\mathbb B} * ,\ \sephetlist {23} {\textit{true}} *)$.

\paragraph{Equality} We may also use dependent types to define inductive propositions, such as equality:

\begin{typedef}{eq}[\forall (A : \Type) \ld A \to A \to \Type]
	\refl ! \forall (A : \Type)\ (x : A) \ld \app {\typename} A x x *
\end{typedef}

\noindent
We use the syntactic sugar $x = y$ for the \texttt{eq} type, meaning $\app {\texttt{eq}} A x y *$ where $A$ is inferred as the type of $x$ and $y$. For example, the equality $2 = 1 + 1$ is a notation for $\app {\texttt{eq}} {\mathbb N} 2 {(1 + 1)} *$.

A useful function utilizing this type is called \emph{transport}. If we have a proof that two values are equal, then we can transport an element of a type indexed by one of them, into one indexed by the other:

\begin{equation*}
	\textit{transport} : \forall (A : \Type)\ (P : A \to \Type)\ (x\ y : A) \ld x = y \to P\ x \to P\ y
\end{equation*}

For example, since we can prove that $\seplist {\mathbb N} {\mathbb B} {\mathbb N} * = \seplist {\mathbb N} {\mathbb B} * \concat [\mathbb N]$, we can transport a value of type $\texttt{HList}\ {\seplist {\mathbb N} {\mathbb B} {\mathbb N} *}$ into one of $\texttt{HList}\ {(\seplist {\mathbb N} {\mathbb B} * \concat [\mathbb N])}$ (in this case, $P$ is the partial type \texttt{HList}). It is often necessary to do this in dependently typed settings, sometimes even multiple times, nested within one another, which may prove troublesome in certain cases. To solve these issues, we will utilize an axiom of dependently typed systems, called Uniqueness of Identity Proofs (UIP)~\cite{uip}; we provide a more detailed discussion on this in \Cref{subsec-uip}.

For convenience, we also introduce \textit{transport\_r}, for transporting in the reverse direction:

\begin{equation*}
	\textit{transport\_r} : \forall (A : \Type)\ (P : A \to \Type)\ (x\ y : A) \ld x = y \to P\ y \to P\ x
\end{equation*}

\paragraph{List membership} Inductively defined propositions may have multiple constructors as well, such as one stating that a value is an element of a list. Its structure mimics that of lists:

\begin{typedef}{In}[\forall (A : \Type) \ld A \to \listof{A} \to \Type]
	\textit{in\_eq} ! \forall x\ xs \ld \app {\typename} x {(x :: xs)} *;
    \textit{in\_cons} ! \forall x\ y\ xs \ld \app {\typename} y {xs} * \to \app {\typename} y {(x :: xs)} *
\end{typedef}

\noindent
The first defining lemma states that the head of the list is included in the list, while the second says that if an element is in the list, then it is in the extended list as well.
Note that elements of this type not only prove the membership but also express the exact position of the value in the list.
We will rely on the above types and concepts in \Cref{sec-syntax,sec-semantics}.

\subsection{Matching logic}\label{subsec-ml}

This subsection introduces the locally nameless representation, the basic syntax and semantics of matching logic~\cite{chen2019mu,chen2023mu}, and describes the theory of definedness, which serves as a standard way to encode equality in matching logic~\cite{chen2019mu}.

The logic is parameterized by a signature, containing sorts, variables, and constant symbols.

\begin{definition}[Signature]
	A matching logic signature is a tuple $(\sepcomma {\Sorts} {\EVar} {\SVar} {\Sigma} *)$, where
	\begin{itemize}
		\item $\Sorts$ is a non-empty set of sorts (denoted with $\sepcomma {s} {s'} {\ldots} *$);
		\item $\EVar = \{\EVar[s]\}_{s \in \Sorts}$ is a countably infinite set of element variables, indexed by sorts, such as $s$ (denoted with $\sepcomma {\ofsortvar{x}{s}} {\ofsortvar{y}{s}} {\ldots} *$);
		\item $\SVar = \{\SVar[s]\}_{s \in \Sorts} $ is a countably infinite set of set variables, indexed by sorts, such as $s$ (denoted with $\sepcomma {\ofsortvar{X}{s}} {\ofsortvar{Y}{s}} {\ldots} *$);
		\item $\Sigma = \{\Sigma_{\sepcomma {s_1} {\dots} {s_n} s *}\}_{\sepcomma {s_1} {\dots} {s_n} s * \in \Sorts}$, where $s_i$ are the parameter sorts and $s$ is the return sort, is a countable set of many-sorted symbols (denoted with $\sepcomma {\sigma} {\sigma'} {\ldots} *$).
	\end{itemize}

	When the details are not relevant, we use $\Sigma$ to refer to the entire signature.
\end{definition}

The formulas of matching logic are called patterns, and fill the role that both predicates and terms do in first-order logic.
As mentioned before, we use a locally nameless representation~\cite{chargueraud2011locally} of the logic, that utilizes traditional named variables when they occur free, and de Bruijn indices~\cite{debruijn1972lambda}, if they are bound by a quantifier.
This choice eliminates the need to reason about $\alpha$-equivalence, and makes it simpler to define capture-avoiding substitutions in a machine-checked formalization.

\begin{definition}[Pattern]\label{def:pattern}
	Given a signature $\Sigma$, the following BNF syntax defines patterns (denoted by $\sepcomma {\varphi_s} {\psi_s} *$, where the subscript indicates the sort of the pattern, i.e. $\varphi_s$ is a pattern of sort $s$):
	\begin{equation*}
		\ofsort{\varphi}{s} \in \textit{Pattern} ::=
		\ofsortvar{\fevar x}{s} \mid
		\ofsortvar{\fsvar X}{s} \mid
		\ofsortvar{\bevar n}{s} \mid
		\ofsortvar{\bsvar N}{s} \mid
		\ofsort{\lnot\varphi}{s} \mid
		\ofsort{\varphi}{s} \land \ofsort{\varphi'}{s} \mid
		\exists[s'] \ld \ofsort{\varphi}{s} \mid
		\mu \ld \ofsort{\varphi}{s} \mid
		\polyapp {\sigma} {\varphi_{s_1}} {\ldots} {\varphi_{s_n}} * \quad \text{if } \sigma \in \Sigma_{\sepcomma {s_1} {\dots} {s_n} s *}
	\end{equation*}

	These constructs are: free element variables, free set variables, bound element variables, bound set variables, negation, conjunction, existential quantifier, least fixed point quantifier, and pattern application. Negation binds the tightest, then conjunction. The scope of the quantifiers extend as far to the right as possible. Note that due to the locally nameless representation, quantifiers do not explicitly bind a variable, instead they use de Bruijn indices. In the $\exists$ binder, the bound variable's sort ($s'$) is explicitly given, while in the case of $\mu$, the sort of the bound variable is the same as the sort of its body. De Bruijn indices represent the number of binders that are in scope between the variable and its corresponding binder. For example, what \cite{chen2019mu} would denote as $\exists \ofsortvar{x}{s} \ld \exists \ofsortvar{y}{s} \ld \ofsortvar{x}{s}$, we write as $\exists[s] \ld \exists[s] \ld \ofsortvar{\bevar 1}{s}$.
We use $\FV{\ofsort{\varphi}{s}}$ to refer to the set of free variables of some pattern $\ofsort{\varphi}{s}$.
\end{definition}

From now on, we omit the sort parameters of patterns, quantifiers, and variables whenever they can be inferred from the context or can be arbitrary. Note that the syntax of patterns above also allows patterns that contain ``dangling'' de Bruijn indices (which are not correctly bound by a quantifier), such as $\exists[s'] \ofsortvar{\bevar 1}{s}$. We call patterns that do not contain such dangling indices \emph{closed}. This notion is sometimes also referred to as ``locally closed'', with ``closed'' also forbidding free variables, however, the semantics of matching logic is equipped to handle free variables, therefore we only need to restrain bound variables in our notion of closedness.

We define capture-avoiding substitutions of patterns in the usual way~\cite{chargueraud2011locally}.

\begin{definition}[Substitution]
	We denote capture-avoiding, many-sorted substitutions as $\subst{\ofsort{\varphi}{s}}{\ofsortvar{x}{s'}}{\ofsort{\psi}{s'}}$ to say that we replace every occurrence of $x$ in $\ofsort{\varphi}{s}$ with $\ofsort{\psi}{s'}$.

	Note that the variable $x$ may be any of the four types of variables (free or bound, element or set variable). In the case of bound variables, only dangling ones can be substituted, and we take into account the quantifiers' ability to increase the index. For example, $\subst{(\exists \ld \bevar 0 \land \bevar 1)}{\bevar 0}{\psi} = \exists \ld (\subst{(\bevar 0 \land \bevar 1)}{\bevar 1}{\psi}) = \exists \ld (\subst{\bevar 0}{\bevar 1}{\psi} \land \subst{\bevar 1}{\bevar 1}{\psi}) = \exists \ld \bevar 0 \land \psi$. This behavior aids in bringing a pattern nearer to closedness, by removing dangling variables.

	Following the idea described in \cite{chargueraud2011locally}, the substituted pattern should be closed. Attempting to substitute with a pattern that is not closed may cause unintended effects, such as capturing a previously dangling variable: $\subst{(\exists \ld \bevar 0 \land \bevar 1)}{\bevar 0}{\bevar 0} = \exists \ld \bevar 0 \land \bevar 0$.
\end{definition}

The logic has a semantics based on pattern matching. Patterns are interpreted as a set of domain elements that match them. Similar to first-order logic, this process is governed by a model, which is defined as follows:

\begin{definition}[Model]
	A ($\Sigma$-)model is a tuple $M = (\sepcomma {\{M_s\}_{s \in \Sorts}} {\{\sigma_M\}_{\sigma \in \Sigma}} *)$, where
\begin{itemize}
	\item $M_s$ is the non-empty carrier set (domain) for each sort;
	\item $\sigma_M : M_{s_1} \times \dots \times M_{s_n} \to \mathcal{P}(M_s)$ serves as the interpretation of the symbol $\sigma \in \Sigma_{\sepcomma {s_1} {\dots} {s_n} s *}$.
\end{itemize}
\end{definition}

Next, we extend the definition of symbol interpretation to model subsets:

\begin{equation*}
	\polyapp {\sigma_M} {A_1} {\dots} {A_n} * \coloneqq \bigcup_{a_i \in A_i} \polyapp {\sigma_M} {a_1} {\dots} {a_n} *
\end{equation*}

In matching logic, element variables evaluate to a single element of the carrier set, while set variables are interpreted as a subset thereof. These mappings are determined by a valuation, denoted with $\rho$.

\begin{definition}[Valuation of variables]
A valuation $\rho$ is a pair of valuations for element and set variables for each $s \in \Sorts$, that is, $\rho : (\EVar[s] \to M_s) \times (\SVar[s] \to \mathcal{P}(M_s))$. We use $\update{\rho}{x}{m}$ to update valuation $\rho$ by mapping the element variable $\ofsortvar{x}{s}$ to the carrier \emph{element} $m \in M_s$. Respectively, we use $\update{\rho}{X}{A}$ to update the valuation of set variables of $\rho$, by mapping the set variable $\ofsortvar{X}{s}$ to the carrier \emph{subset} $A \subseteq M_s$.
\end{definition}

With this concept, we define the interpretation of patterns (denoted by $\bar \rho$). This function has a simplified type, $\textit{Pattern} \to \mathcal P(M_s)$. We note that (dangling) bound variables are interpreted as $\emptyset$ similar to our previous work~\cite{bereczky2022mechanizing}. In practice, we do not construct ill-formed patterns (which contain dangling de Bruijn indices).

\begin{definition}[Matching logic semantics] Given a matching logic signature $\Sigma$, model $M$, and a valuation $\rho$, we define pattern interpretation (mapping patterns of sort $s$ to subsets of $M_s$) the following way:

\begin{gather*}
	\bar \rho (\ofsortvar{\fevar x}{s}) = \{\rho (x)\} \qquad\qquad 
	\bar \rho (\ofsortvar{\fsvar X}{s}) = \rho (X) \qquad\qquad 
	\bar \rho (\ofsortvar{\bevar n}{s}) = \emptyset \qquad\qquad 
	\bar \rho (\ofsortvar{\bsvar N}{s}) = \emptyset \\
	\bar \rho (\ofsort{\varphi}{s} \land \ofsort{\varphi'}{s}) = \bar \rho (\ofsort{\varphi}{s}) \cap \bar \rho (\ofsort{\varphi'}{s}) \qquad\qquad
	\bar \rho (\lnot \ofsort{\varphi}{s}) = M_s \setminus \bar \rho (\ofsort{\varphi}{s}) \\
	\bar \rho (\exists[s'] \ld \ofsort{\varphi}{s}) = \bigcup_{m \in M_{s'}} \overline{\update{\rho}{x}{m}} (\subst{\ofsort{\varphi}{s}}{\ofsortvar{\bevar 0}{s'}}{\ofsortvar{x}{s'}}), \text{where } x \notin \FV{\ofsort{\varphi}{s}} \\
  \bar \rho (\mu \ld \ofsort{\varphi}{s}) = \lfp \mathcal{F}^\rho_{\ofsort{\varphi}{s}, \ofsortvar{X}{s}}, \text{where } \mathcal{F}^\rho_{\ofsort{\varphi}{s}, \ofsortvar{X}{s}}(A) = \overline{\update{\rho}{X}{A}} (\subst{\ofsort{\varphi}{s}}{\ofsortvar{\bsvar 0}{s}}{\ofsortvar{X}{s}}) \text{ and } X \notin \FV{\ofsort{\varphi}{s}}\\
	\bar \rho (\polyapp {\sigma} {\varphi_{s_1}} {\dots} {\varphi_{s_n}} *) = \polyapp {\sigma_M} {\bar\rho(\varphi_{s_1})} {\dots} {\bar\rho(\varphi_{s_n})} *, \text{where } \sigma \in \Sigma_{\sepcomma {s_1} {\dots} {s_n} s *}
\end{gather*}
We use $\lfp \mathcal{F}$ to denote the least fixpoint of a monotone function $\mathcal{F}$, and define it as an intersection of pre-fixpoints as in~\cite{chen2019mu}.\footnote{Note that $\mathcal{F}^\rho_{\ofsort{\varphi}{s}}$ is only monotone when $\ofsort{\varphi}{s}$ is positive~\cite{chen2019mu}, which is crucial when proving the soundness of a logical deduction system. However, in this paper, we do not define syntactic proofs.}
\end{definition}

The logic also has a proven-sound Hilbert-style proof system, however, this paper focuses on semantic applications, therefore we will not discuss it here, and we refer the interested reader to~\cite{bereczky2022mechanizing,chen2019mu}.

The logic may be ``extended'' with theories consisting of new sorts, symbols, notations, and a set of axioms.\footnote{While in first-order logic, the term \emph{theory} is commonly used to refer only to a set of axioms, in matching logic it is used to mean the entire specification, which includes symbols and notations.} 
One commonly used theory is called definedness~\cite{chen2019mu}. It introduces one new symbol to the signature, denoted with $\defined{\_}[s][s'] \in \Sigma_{s, s'}$. The notation $\defined{\ofsort{\varphi}{s}}[s][s']$ is used to write $\app {\defined{\_}[s][s']} {\ofsort{\varphi}{s}} *$.
The definedness theory consists a single axiom, stating for every $x$ that $\defined{\ofsortvar{\fevar x}{s}}[s][s']$. 
Furthermore, the semantics of the definedness symbol can be expressed with the following equation:

\begin{equation*}
	\bar \rho (\defined{\ofsort{\varphi}{s}}[s][s']) = \begin{cases}
		\emptyset & \text{if } \bar \rho (\ofsort{\varphi}{s}) = \emptyset \\
		M_{s'} & \text{otherwise}
	\end{cases}
\end{equation*}
\noindent
Based on definedness, one can derive equality in a standard way; we refer to~\Cref{subsec-extpat} and~\cite{chen2019mu}, because these details are beyond the scope of this paper.

\section{Related work}\label{sec-work}

\subsection{Existing formalizations of matching logic}\label{subsec-aml-form}

Logical frameworks such as first-order logic, temporal logic~\cite{8133459}, separation logic~\cite{IrisProofMode}, or indeed matching logic are often embedded into proof assistants. This allows checking the correctness of any formalization or proof using a computer which greatly increases the trustworthiness of any such system. Matching logic has many such embeddings, including one in Metamath~\cite{ml-in-mm} (used for generating proof certificates~\cite{lin2023generating}, however, lacking semantic reasoning), one in LEAN4~\cite{aml-lean}, and several in \Coq{}, such as a general-purpose shallow embedding~\cite{aml-in-coq-traian}, a shallow embedding aimed at unification~\cite{aml-in-coq-unif}, and a deep embedding~\cite{aml-harp} which also implements a proof-mode for matching logic based on a sequent calculus~\cite{tusil}. All these formalizations express the ``applicative'' variant of matching logic, which features binary applications and \emph{unsorted} syntax.

\paragraph{Applicative matching logic in Coq}

The closest related result to the efforts in this work is the Coq implementation of the applicative, unsorted variant of matching logic~\cite{aml-harp}. While it would be possible to introduce sorts into this using a well-sortedness predicate attached to patterns, proving and maintaining these through changes made to the patterns is cumbersome. The semantics also has no support for sorts, and certain elements of the carrier need to be designated as belonging to a certain sort separately as well. We aim to remedy this by introducing sorts at the type level.

It also treats applications as binary operations and symbols as a separate construct. Due to the lack of sorts, the symbols also lack a built-in sort signature. Introducing sorts proves more difficult with binary applications. The carrier must also contain elements for partial applications. Another feature of this implementation is the well-formedness predicate. Since this formalization also uses the locally nameless representation of variables, patterns are not inherently well-formed. There can be dangling bound variables that need to be forbidden by a closedness judgement. We will encode all this at the type level.

\subsection{Term algebras with substitutions}\label{subsec-subcalc}

There are well-known descriptions of term algebras that can be used to describe programming languages~\cite{pfpl}. Such descriptions feature term types which are indexed by their object-theoretic types, as well as a variable context. This context is used to define closures, such as lambda functions. The variables corresponding to this context are a version of naturals indexed by their type and context. The algebraic theory also features substitutions, with a type describing their effect on a context. These substitutions are constructed similarly to variables with the base case containing the term to be substituted. There is also a special substitution that serves only to increment variables.

Note however that the working of substitutions is implemented using equations bundled with the algebra. This is reminiscent of quotient inductive types which lack proper support in current proof assistants. In the next section we describe our formalization of matching logic syntax similar to a term algebra; in particular, \Cref{subsec-subs} presents a modified, computable version of substitutions, adapting the core ideas from the substitution calculus.

\section{Dependently typed syntax}\label{sec-syntax}

In this section we describe the dependently typed syntax of matching logic.
Before we begin defining the syntax, we first need to establish the signature that serves as its foundation.

\subsection{Signature}\label{subsec-signature}

The signature includes a set of sorts, that essentially serves as the object-level representation of types. Additionally, we have two indexed sets for element and set variables. These sets are indexed by the sorts of the signature, ensuring that each sort has its own sets for the two types of variables. This allows variables of different sorts to be distinct types, while also preventing the same variable from representing elements of different sorts. Finally, the signature incorporates symbols, along with mappings that associate each symbol with its parameter and return sorts. The formal definition is shown in \Cref{fig-sig}.

\begin{figure*}[htb]
	\begin{typedef}{\Sorts}*[\Type]
		\EVar ! \Sorts \to \Type;
		\SVar ! \Sorts \to \Type;
		\Sigma ! \Type;
		\argsorts ! \Sigma \to \listof{\Sorts};
		\retsorts ! \Sigma \to \Sorts
	\end{typedef}
	\caption{Dependently typed signature of matching logic.}
	\label{fig-sig}
\end{figure*}

In the following subsections we showcase the steps of integrating sortedness and closedness into the basic syntax of the logic, then we extend the final version of the syntax with definedness, and finally we discuss substitutions.

\subsection{Sortedness}\label{subsec-sortedness}

First, we examine how sorts are represented in the dependently typed setting. We define patterns to be indexed by their sorts in \Cref{fig-sorted}. In the following figures, we reuse the notations for patterns introduced in \Cref{def:pattern}, and $\placeholder$ is used to denote the ``holes'' where the concrete parameters are placed. Note that for application patterns, we use $\cdot$ as the constructor, which separates symbols from their arguments represented as (heterogeneous) lists.

\begin{figure*}[htb]
	\begin{typedef}{Pattern}[\Sorts \to \Type]
		\fevar \placeholder ! \forall (s : \Sorts) \ld (\app {\EVar} {s} *) \to \sortedpattern{s};
		\fsvar \placeholder ! \forall (s : \Sorts) \ld (\app {\SVar} {s} *) \to \sortedpattern{s};
		\bevar \placeholder ! \forall (s : \Sorts) \ld \mathbb N \to \sortedpattern{s};
		\bsvar \placeholder ! \forall (s : \Sorts) \ld \mathbb N \to \sortedpattern{s};
		\syapp \placeholder \placeholder ! \forall (\sigma : \Sigma) \ld \app {\texttt{HList}} {(\map{\texttt{Pattern}}{(\app {\argsorts} {\sigma} *)})} * \to \sortedpattern{(\app {\retsorts} {\sigma} *)};
		\lnot \placeholder ! \forall (s : \Sorts) \ld \sortedpattern{s} \to \sortedpattern{s};
		\placeholder \land \placeholder ! \forall (s : \Sorts) \ld \sortedpattern{s} \to \sortedpattern{s} \to \sortedpattern{s};
		\exists \ld \placeholder ! \forall (s : \Sorts) \ld \sortedpattern{s} \to \sortedpattern{s};
		\mu \ld \placeholder ! \forall (s : \Sorts) \ld \sortedpattern{s} \to \sortedpattern{s}
	\end{typedef}
	\caption{Definition of sorted patterns. When using these patterns, we treat $s$ as an implicit parameter. The $\map {\_} {\_}$ function is the one introduced in \Cref{subsec-metatypes}.}
	\label{fig-sorted}
\end{figure*}

Free (set and element) variables are indexed by their respective sorts in the signature. Bound variables can be of any sort (we will restrict them in \Cref{subsec-closed-sortedness}).
Most of the recursive constructors inherit the sorts of their arguments if they are the same (e.g., $\lnot \varphi$ or $\exists \ld \varphi$ have the same sort as $\varphi$, while $\varphi_1 \land \varphi_2$ requires that both $\varphi_1$ and $\varphi_2$ have the same sort and the conjunction also adapts that sort). For example, a pattern such as $\bevar 0 \land \bsvar 1$ can be constructed only if the two sides of the conjunction have the same sort; however, as this can be any sort, it cannot be inferred without any binding context.

Pattern application is an uncurried variadic constructor of patterns, where the final sort does not correspond directly to any of the argument sorts, instead, it is determined by the signature of the symbol. We express this with heterogeneous lists (described in \Cref{subsec-dependenttt}), where the elements are patterns with sorts determined by the symbol, and the overall pattern's sort matches the return sort of the symbol. This means that, unlike the previous example, once the symbol is specified, all sorts within this pattern are known.
This behavior facilitates the inference of sorts for other parts of a pattern containing applications. For example, given sorts $\nat$ and $\bool$, and a symbol $isZero$ with the signature $\nat \to \bool$, then in the pattern $(\syapp {isZero} {\sephetlist {\varphi_1} *}) \land \varphi_2$, $\varphi_1$ is required to have sort $\nat$, while $\varphi_2$ must be of sort $\bool$.

Finally we remark that in this intermediary representation, the types of the $\exists$ and $\mu$ binders are the same, as we are only indexing by the sort of the overall pattern. In then next subsection, there will be a small difference, and in the subsection afterwards, the differences outlined in \Cref{subsec-ml} will become visible with the introduction of a proper typing environment.

\subsection{Closedness}\label{subsec-closedness}

In this subsection, we describe how to enforce the closedness of patterns using the type system. First, we showcase this through unsorted patterns, as incorporating sorts would significantly complicate this presentation, and we aim to keep this introduction to closedness clear and concise. In the following subsection, we will reintroduce the concept of sortedness to closed patterns.

A pattern is considered closed if it contains no dangling bound variables. This concept can be extended to ``being closed up to a certain index'', meaning the pattern contains no dangling bound variables with an index greater than or equal to the one specified. For instance, the pattern $\bevar 0 \land \bevar 2$ is closed up to index 3, as the highest dangling index is 2. Note that this pattern is also closed up to any index greater than 3, since 2 is less than any such number. Thus, this index serves as an upper bound. A pattern is considered closed if it is closed up to index 0, (as there are no indices lower than 0, this means that the pattern may not contain any dangling variables).

We represent the index of closedness using two natural numbers as type indices for patterns, one for each quantifier ($\exists$ and $\mu$). These serve as the ``up to'' bounds mentioned earlier. The definition of closed patterns is showcased in \Cref{fig-closed}.

\begin{figure*}[htb]
	\begin{typedef}{Pattern}[\mathbb N \to \mathbb N \to \Type]
		\fevar \placeholder ! \forall (s : \Sorts)\ (n\ m : \mathbb N) \ld (\app {\EVar} {s} *) \to \closedpattern{n}{m};
		\fsvar \placeholder ! \forall (s : \Sorts)\ (n\ m : \mathbb N) \ld (\app {\SVar} {s} *) \to \closedpattern{n}{m};
		\bevar \placeholder ! \forall (n\ m\ k : \mathbb N) \ld \closedpattern{(k + n)}{m};
		\bsvar \placeholder ! \forall (n\ m\ k : \mathbb N) \ld \closedpattern{n}{(k + m)};
		\syapp \placeholder \placeholder ! \forall (n\ m : \mathbb N) \ld \Sigma \to \listof{(\closedpattern{n}{m})} \to \closedpattern{n}{m};
		\lnot \placeholder ! \forall (n\ m : \mathbb N) \ld \closedpattern{n}{m} \to \closedpattern{n}{m};
		\placeholder \land \placeholder ! \forall (n\ m : \mathbb N) \ld \closedpattern{n}{m} \to \closedpattern{n}{m} \to \closedpattern{n}{m};
		\exists \ld \placeholder ! \forall (n\ m : \mathbb N) \ld \closedpattern{(\app S n *)}{m} \to \closedpattern{n}{m};
		\mu \ld \placeholder ! \forall (n\ m : \mathbb N) \ld \closedpattern{n}{(\app S m *)} \to \closedpattern{n}{m}
	\end{typedef}
	\caption{Definition of closed patterns. In this definition, $s$, $n$, and $m$ are treated as implicit parameters.}
	\label{fig-closed}
\end{figure*}

Most base cases can have arbitrary closedness indices, while recursive constructors propagate their arguments' indices, similar to sortedness. The special base cases here are bound variables, as they may only be considered closed up to any index that is greater than their own.

In the case of applications, each argument in the list needs to have the same closedness. Since the symbol does not enforce a fixed value, this can also be arbitrary. Note that in this case, the list does not need to be heterogeneous.

The quantifiers increase the index of their subpatterns by one. For instance, if $\exists \ld \varphi$ is closed up to 3 for the $\exists$ binders, then $\varphi$ must be closed up to 4. This allows the additional variable bound by the quantifier to appear in the subpattern. Existential quantification only influences the first index, while fixpoint binders increase the second one.

Note that even quantifiers do not fully constrain the index of their subpatterns---they only require it to be incremented by one. For example, in the pattern $\mu \ld \fevar x$, the closedness index of $\fevar x$ can be any value above 0. Similarly, for bound variables, a pattern such as $\fevar x \land \bsvar 3$ is only required to be closed up to at least 4 indices for fixed points. The exact indices can be determined by type inference.

In the case of recursive constructors, we could have chosen any index greater than or equal to the maximum of their arguments' indices. This is not necessary, however, since if the environment requires a specific closedness for a pattern, this requirement can be propagated down to the construction of the base cases. This eliminates the need to have a metatheoretical function appearing in the type, which might not simplify or be inferable in every case.

For example, if we had a pattern where one of the closedness indices is $k + \app {\max} n 0 *$, and we can prove that this is equal to $k + n$; however, as these indices are not definitionally equal, we must prove the equality separately, and utilize a transport (introduced in \Cref{subsec-dependenttt}) to change the type (which might cause further complications when this pattern is used). Similarly, if we require a pattern with a fixed index, let us say $9$, we cannot immediately use these constructors, as $9$ is not unifiable with $k + \app {\max} n m *$, if $k$, $n$, or $m$ are not known. We are required to either provide every uninferable implicit argument so the expression would simplify to $9$, or use the previous method of transporting using some proof to make them equal.

\subsection{Connecting closedness and sortedness}\label{subsec-closed-sortedness}

In this subsection, we combine the results from the previous two subsections and explain how closedness may be encoded in the sorted setting.
With the introduction of sorts, we can generalize the natural numbers representing closedness into a \emph{sorting environment}, represented as a list of sorts, similar to the substitution calculus discussed in \Cref{subsec-subcalc}. 
These sorting environments encode the sorts of all dangling bound variables in the pattern. The indices in the list correspond to the indices of the variables. Specifically, the \nth[i] element of the list describes the sort of the variable with the \nth[i] index. The length of the list indicates the closedness of the pattern, and if both lists are empty, it is considered closed. This definition of patterns can be found in \Cref{fig-closed-sorted}.

\begin{figure*}[htb]
	\begin{typedef}{Pattern}[\listof{\Sorts} \to \listof{\Sorts} \to \Sorts \to \Type]
		\fevar \placeholder ! \forall (s : \Sorts)\ (ex\ mu : \listof \Sorts) \ld (\app {\EVar} {s} *) \to \sortedclosedpattern{s}{ex}{mu};
		\fsvar \placeholder ! \forall (s : \Sorts)\ (ex\ mu : \listof \Sorts) \ld (\app {\SVar} {s} *) \to \sortedclosedpattern{s}{ex}{mu};
		\bevar \placeholder ! \forall (s : \Sorts)\ (ex\ mu : \listof \Sorts) \ld \app {\texttt{In}} s {ex} * \to \sortedclosedpattern{s}{ex}{mu};
		\bsvar \placeholder ! \forall (s : \Sorts)\ (ex\ mu : \listof \Sorts) \ld \app {\texttt{In}} s {mu} * \to \sortedclosedpattern{s}{ex}{mu};
		\syapp \placeholder \placeholder ! \forall (ex\ mu : \listof \Sorts)\ (\sigma : \Sigma) \ld \app {\texttt{HList}} {(\map{(\closedpattern{ex}{mu})}{(\app {\argsorts} {\sigma} *)})} * \to ! \sortedclosedpattern{(\app {\retsorts} {\sigma} *)}{ex}{mu};
		\lnot \placeholder ! \forall (s : \Sorts)\ (ex\ mu : \listof \Sorts) \ld \sortedclosedpattern{s}{ex}{mu} \to \sortedclosedpattern{s}{ex}{mu};
		\placeholder \land \placeholder ! \forall (s : \Sorts)\ (ex\ mu : \listof \Sorts) \ld \sortedclosedpattern{s}{ex}{mu} \to \sortedclosedpattern{s}{ex}{mu} \to ! \sortedclosedpattern{s}{ex}{mu};
		\exists[\placeholder] \ld \placeholder ! \forall (s\ s' : \Sorts)\ (ex\ mu : \listof \Sorts) \ld \sortedclosedpattern{s}{(s' :: ex)}{mu} \to \sortedclosedpattern{s}{ex}{mu};
		\mu \ld \placeholder ! \forall (s : \Sorts)\ (ex\ mu : \listof \Sorts)\ \sortedclosedpattern{s}{ex}{(s :: mu)} \to \sortedclosedpattern{s}{ex}{mu}
	\end{typedef}
	\caption{Definition of sorted closed patterns. $s$, $ex$, and $mu$ are implicit parameters, however, $s'$ is not.}
	\label{fig-closed-sorted}
\end{figure*}

The lists and the sort behave similarly to those described in the previous subsections, with most recursive cases inheriting their subpatterns' environments and sorts. The primary difference is with quantifiers and bound variables.

The existential quantifier extends the environment of its subpattern with a new arbitrary sort (its parameter is of type $\sortedclosedpattern{s}{(s' :: ex)}{mu}$, for some parameter sort $s'$). This corresponds to the newly bound variable, and the fact that the remaining sorts are pushed back demonstrates how all indices are shifted up by one under the quantifier. Fixed point binders are constructed similarly; however, they extend their subpattern's environment with their own sort (its type is $\sortedclosedpattern{s}{ex}{(s :: mu)}$, where $s$ is the sort of the resulting pattern as well) rather than an arbitrary one (as also discussed in \Cref{subsec-ml}).

While free variables are assigned sorts by the signature, bound variables, previously represented by natural numbers (\Cref{subsec-sortedness}), lack this explicit typing information. However, we can remedy this by assigning the \nth[i] sort in the environment for element variables to the bound variable $\bevar i$ (and the same idea applies for set variables). To avoid indexing out of bounds, we can utilize the \texttt{In} judgement (defined in \Cref{subsec-dependenttt}) to represent bound variables instead of natural numbers. We note that there is still a direct correspondence between values of type \texttt{In} and natural numbers: the constructor \textit{in\_eq} can be seen as the index $O$, while \textit{in\_cons} represents $S$, the successor. The bound variable $\bevar{\app {\textit{in\_eq}} {s} {\textit{ex}} *}$ represents $\bevar 0$ and a proof that the element of index $0$ in the environment $(\sep {::} {s} {\textit{ex}} *)$ is $s$. In a similar way, $\bevar{\app {\textit{in\_cons}} {s_1} {s_2} {(s_2 :: \textit{ex})} {(\textit{in\_eq}\ {s_2}\ {\textit{ex}})} *}$ represents bound variable $\bevar 1$ and the proof that the element of index $1$ in the environment $(\sep {::} {s_1} {s_2} {\textit{ex}} *)$ is $s_2$. For readability, in the rest of the paper, we use natural numbers to denote bound variables, but we highlight that these numbers correspond to values of the judgement \texttt{In} as presented here.

Type inference is also possible in this setting. For example, given the same $isZero$ symbol as above, with type $\nat \to \bool$, the pattern $\exists_{s} \ld \syapp {isZero} {\sephetlist {\bevar 1} *}$ has type $\sortedclosedpattern{\bool}{(\sep {::} s \nat {ex} *)}{mu}$, for any sort $s$ and environments $ex$ and $mu$, meaning that the only requirement is the existence of at least two dangling bound element variables, with the second of these being of sort $\nat$, and that the overall pattern's sort is $\bool$. Beyond these conditions, both contexts may be extended arbitrarily.

Introducing the sorting environment is a generalization, as the unsorted setting can be viewed as having a single sort, where every pattern has this sort, and the natural number type indices of \Cref{fig-closed} correspond to lists containing as many instances of this single sort as the number (e.g., $3 \equiv \seplist {\star} {\star} {\star} *$, where $\star$ denotes the single sort). For example the a type from the previous subsection, such as $\closedpattern{3}{2}$ is isomorphic to $\sortedclosedpattern{\star}{\seplist {\star} {\star} {\star} *}{\seplist {\star} {\star} *}$ in this subsection.

\subsection{Definedness}\label{subsec-def-in-syntax}

For the rest of this paper, we will utilize a slightly modified syntax of the logic, in which the theory of definedness is deeply embedded. Specifically, the description of patterns from \Cref{subsec-ml} is extended to include the definedness symbol, as shown in \Cref{subsec-ml}. We extend the definition of \texttt{Pattern} shown in \Cref{subsec-closed-sortedness} with one more constructor ($s$, $ex$, and $mu$ are implicit parameters):

\begin{equation*}
	\defined{\placeholder}[][\placeholder] : \forall (s\ s' : \Sorts)\ (ex\ mu : \listof \Sorts) \ld \sortedclosedpattern{s}{ex}{mu} \to \sortedclosedpattern{s'}{ex}{mu}
\end{equation*}

An important distinction between this constructor and the others is that it may have an arbitrary sort regardless of the sort of its argument. This is because it forms a predicate pattern, whose interpretation is either the entire set or the empty set, depending on whether its argument contains any elements. However, the sort of the resulting set does not need to be linked to the sort of the argument's set.

This also implies that in a pattern such as $\defined{\bsvar 0}[][s']$, the sort of the pattern, $s'$, must be explicitly stated, as it may not be inferred from the argument's sort, since it is completely independent of it.

\subsection{Substitutions}\label{subsec-subs}

A fundamental operation on patterns necessary for defining the semantics is substitution. There are two primary cases of substitution: one for free variables and another for bound variables. Each of these may be further subdivided into element and set variables, however, in both main cases, these can be defined analogously.

An important transformation on patterns that we need to implement for both types of substitutions is the extension of either environment. The type of this is

\begin{align*}
&\forall ex\ ex'\ ex''\ mu\ mu'\ mu''\ s \ld \\
&\qquad \sortedclosedpattern {s} {(ex \concat ex'')} {(mu \concat mu'')} \to \sortedclosedpattern {s} {(ex \concat ex' \concat ex'')} {(mu \concat mu' \concat mu'')}
\end{align*}
\noindent
where $ex \concat ex''$ and $mu \concat mu''$ are the original environments to be extended by $ex'$ and $mu'$, split at the point where the insertion needs to happen. This is defined recursively: for most constructors, we simply rebuild that pattern with the new, extended indices. The only exceptions are bound variables, where we must also increment the variables that come after the extension. This ``incrementation'' corresponds to the following lemma\footnote{\label{coqfootnote}We note that in the \Coq{} implementation we need to restate these lemmas, since we need them to be computable, which the built in lemmas usually are not. We also remark that the \texttt{In} type introduced in this paper is not the same as the one in the standard library of \Coq{}.} on the \texttt{In} type (see \Cref{subsec-dependenttt}):

\begin{equation*}
	\textit{In\_concat} : \forall (A : Type)\ (s : A)\ (xs\ ys\ zs : \listof A) \ld \app {\texttt{In}} s {(xs \concat zs)} * \to \app {\texttt{In}} s {(xs \concat ys \concat zs)} *
\end{equation*}

This expresses that if a value is in a list, it will also be in the same list, extended anywhere. In the case of quantifiers, we are required to perform two transports using the following lemma\cref{coqfootnote} about the associativity of list operations:

\begin{equation*}
	\textit{List\_assoc} : \forall (A : \Type)\ (xs\ ys : \listof A)\ (a : A) \ld a :: (xs \concat ys) = (a :: xs) \concat ys
\end{equation*}

For an example usage of this function, consider the pattern $\bevar 1 \land \bevar 2$ of type $\sortedclosedpattern{s'}{\seplist {s_0} {s'} {s'} *}{[]}$. Extending its existential environment after the first $s'$ by $\seplist {s'_2} {s'_3} *$ results in $\sortedclosedpattern{s'}{\seplist {s_0} {s'} {s'_2} {s'_3} {s'} *}{[]}$. The new pattern corresponding to this type is $\bevar 1 \land \bevar 4$. Note that the first variable remains unchanged, as the extension occurs after it, whereas the second variable is properly incremented. However, the implicit arguments of both variables are modified to reflect the extended environment in which they now reside. This operation is required to express capture avoidance of substitutions by increasing dangling indices of the substituted patterns.

\paragraph{Free variable substitution} Since free variables are only indexed by sorts, this operation is similar to the non-dependent version. It is important to allow the pattern in which we perform substitution to have a larger context than the one being substituted, enabling us to execute the operation under quantifiers as well. This results in the type
\begin{equation*}
\forall ex\ ex'\ mu\ s\ s' \ld \sortedclosedpattern {s'} {ex'} {mu} \to (\app {\EVar} {s'} *) \to \sortedclosedpattern {s} {(ex \concat ex')} {mu} \to \sortedclosedpattern {s} {(ex \concat ex')} {mu}
\end{equation*}
\noindent
for element variable substitution, and a similar one for set variables, where we do the same changes to the other context. Here, the first pattern is the one being substituted, and the second is the one in which the substitution is being performed. Most of the base cases remain unaffected by the operation, while the recursive cases perform the substitution on their subpatterns, with quantifiers extending the environment as described above, if necessary.

In the cases involving the constructors of free variables, two equality checks are necessary. First, we decide whether the sorts of the variable to be replaced and the target variable are equal. If they are, we then perform a transport based on this equality on one of the variables (to unify their sorts) and compare it with the other variable. If they are also equal, we proceed to replace it with the target pattern, extending its environment as needed. If either check fails, the variables are considered unequal, and we need not modify it. This transport causes no issues in practice as it can be discharged automatically or by utilizing UIP (described in \Cref{subsec-dependenttt} and \Cref{subsec-uip}).

We present 3 representative cases of this function, and refer to the Coq implementation~\cite{kore-ml} for the rest of them. We also omit the parameters that are not relevant and may be inferred from the patterns.
\begin{align*}
	\app {\textit{fevar\_subst}} {\psi} {x} {\bevar n} * &= \bevar n \\
	\app {\textit{fevar\_subst}} {\psi} {x} {(\varphi_1 \land \varphi_2)} * &= \app {\textit{fevar\_subst}} {\psi} {x} {\varphi_1} * \land \app {\textit{fevar\_subst}} {\psi} {x} {\varphi_2} * \\
	\app {\textit{fevar\_subst}} {\psi} {(\ofsortvar{x}{s'})} {(\ofsortvar{\fevar{y}}{s})} * &= \begin{cases}
		\app {\textit{transport}} H {\psi} * & \text{if } (H : s' = s) \land \app {\textit{transport}} H {(\ofsortvar{x}{s'})} * = \ofsortvar{y}{s} \\
		\ofsortvar{\fevar{y}}{s} & \text{otherwise}
	\end{cases}
\end{align*}

\paragraph{Bound variable substitution} In order to define the substitution of bound variables in patterns, we first need to specify how indices are to be substituted. The operation has the type 
\begin{equation*}
\textit{In\_subst} : \forall s\ s'\ ex\ ex'\ mu \ld\allowbreak \app {\texttt{In}} s {(ex \concat s' :: ex')} * \to \sortedclosedpattern {s'} {ex'} {mu} \to \sortedclosedpattern {s} {(ex \concat ex')} {mu}
\end{equation*}
\noindent
for the existential context, and can be similarly typed for the fixed point one. We define this by recursion on the index. There are four possible cases, depending on the current index and the index to be substituted:

\begin{itemize}
    \item If both are zero, we have reached the end of the substitution, and we may simply return with the replacement pattern. Note that in this case, both the sort and environment of the variable index and the substituted pattern are equal, highlighting that the indices function like placeholders for substitution.
    \item If the current index is zero, but the index to be substituted is higher, then we also reached the last iteration, and it has become clear that we are attempting to substitute an index higher than what we have. In this case there is nothing to do, and we can simply return $\bevar 0$.
    \item If the current index is not zero (i.e. $\app S n *$ for some $n$ index), but the target index is zero, then in this base case, we see that the substitution is for a lower index than what we had. This is similar to the previous case, however, since this time the operation will make the environment of the pattern smaller, we must decrement the current index by one, returning $\bevar n$.
    \item Finally, if neither index is zero, we need to perform this operation recursively, decreasing both indices by one. It is important to note that we do not want to lose the first variable of the environment or any of the $S$ constructors on the returned index if the substitution cannot be performed. Therefore, after the recursive call, we must reextend the environment of the returned pattern by the first variable, using the method described above.
\end{itemize}

\noindent
With the above function established, we can define bound variable substitution over patterns with type
\begin{equation*}
\forall ex\ ex'\ mu\ s\ s' \ld \sortedclosedpattern {s'} {ex'} {mu} \to \sortedclosedpattern {s} {(ex \concat s' :: ex')} {mu} \to \sortedclosedpattern {s} {(ex \concat ex')} {mu}
\end{equation*}
Like before, set variable substitution may be typed similarly. We also highlight that it is not necessary to provide the index to be substituted explicitly, as it can be inferred from the environments and sorts of the patterns; therefore, only the pattern to be substituted and the one to substitute in are necessary. In the definition, for the bound variables themselves, we may defer to the behavior described earlier. Otherwise, this process is similar to free variable substitution, where we reconstruct the pattern, and use recursive calls and environment extension as needed.

It is also important to note that, similar to free variable substitution, the environment of the pattern to be substituted may be extended compared to the pattern in which we do substitution. However, unlike previous operations, it must also be extended by the sort of the replacement pattern, since this substitution also removes that sort from the environment.

We highlight some cases of the definition. Like before, we omit implicit parameters and analogously definable cases:

\begin{align*}
	\app {\textit{bevar\_subst}} {\psi} {\fevar x} * &= \fevar x \\
	\app {\textit{bevar\_subst}} {\psi} {\bevar{n}} * &= \app {\textit{In\_subst}} n \psi * \\
	\app {\textit{bevar\_subst}} {\psi} {(\exists[s''] \ld \varphi)} * &= \app {\textit{transport\_r}} {\textit{List\_assoc}} {(\app {\textit{bevar\_subst}} {\psi} {(\app {\textit{transport}} {\textit{List\_assoc}} {\varphi} *)} *)} *
\end{align*}

Note that in the third case, $\varphi$ is of type $\sortedclosedpattern {s} {(s'' :: ex \concat s' :: ex')} {mu}$, which is transported to $\sortedclosedpattern {s} {((s'' :: ex) \concat s' :: ex')} {mu}$. After applying the substitution to this pattern, the result is of type $\sortedclosedpattern {s} {((s'' :: ex) \concat ex')} {mu}$, where we use the outer transport to get rid of the parentheses.

\section{Dependently typed semantics}\label{sec-semantics}

In this section, we define the semantics for the syntax introduced in the previous section.

\subsection{Dependent models}\label{subsec-depmodels}

The model must include a carrier set whose subsets the patterns are mapped to. We have chosen to index these by the sorts.

This approach allows the sorts to directly correspond to types in the metatheory (e.g., $\nat$ to natural numbers, or a sort whose domain is described using a least fixed point to a custom inductive type). This can be achieved either with a single monolithic inductive type, that copies the structure of or directly embeds these corresponding types, or via a function that maps to smaller, independently defined types.

This latter approach is powerful because the definition simplifies directly to the intended type, whereas in the former, it may be necessary to wrap and unwrap constructors and to refute cases created with the invalid constructors (e.g., $\nat$-carriers cannot have been created using the $\bool$-carrier's constructor). The indexed carrier also enables a simpler definition for the interpretation of symbols and their applications, since it only requires definitions for well-sorted cases. Finally, it also makes it much simpler to express the requirement that the carrier must be inhabited for every sort.

When defining the interpretation of symbols and their applications, we employ a heterogeneous list for the arguments, similar to the syntax, that contains elements of the carriers of the argument sorts, and return a set containing elements of the carrier of the return sort, all based on the signature of the given symbol.

Based on this, we can also define the extended application function, where the key difference is that the heterogeneous list of arguments now contains a set of elements from the carriers of the given sorts, and the result is computed from all possible combinations of these sets' elements. However, this simple modification significantly increases the complexity of reasoning about applications. We will address this issue for a common special case in \Cref{subsec-satisfiability}.

\subsection{Valuations}\label{subsec-valuations}

Since both variables and the carrier are indexed by sorts, variable valuations can take advantage of this by mapping variables of a particular sort only to the carrier of that sort, or to a set thereof in the case of set variables. This approach simplifies working with free variables during pattern evaluation.

However, sortedness also introduces additional complexity in certain definitions, such as updating a specific variable within the valuation. Similar to the process outlined in \Cref{subsec-subs} for free variable substitutions, it is necessary to perform a double equality check, verifying that both the sort of the variables and, after a transport, the variables are equal.
The extended valuation function may be defined analogously to the formal definition shown in \Cref{subsec-ml}.

\subsection{Dependently typed theories and satisfaction}\label{subsec-satisfiability}

To prove that a given theory satisfies a particular model, it is necessary to first define dependently typed theories and model satisfaction in this context.

The axioms of a theory are represented as a set of dependent pairs, each consisting of a sort and a pattern of that sort. This structure allows us to specify axioms for a single sort or for all of them. This is especially important for axioms involving definedness, as they may have arbitrary sorts, and we typically expect such axioms to hold for any sort.

A model is considered to satisfy a theory with axioms of this form if, for each axiom in the set, the extended valuation maps it to the entire carrier set of its specified sort.

Since the valuation transforms every pattern into a set expression, satisfiability primarily involves set-theoretic reasoning. Much of this can often be simplified or solved automatically. However, simplifying the extended interpretation of the application function can be challenging, as it requires reasoning about a set constructed from every combination of a list of sets.

Consider a theory for example, containing two sorts, $\bool$ and $\nat$. Their domains may be described with the patterns $\textit{false} \lor \textit{true}$ and $\mu \ld O \lor \app S {\bsvar 0} *$, and we may define various operations for them, such as the aforementioned $\textit{isZero}$, with signature $\nat \to \bool$, that evaluates to $\textit{true}$ if applied to $O$ and $\textit{false}$ otherwise, as well as various well-known boolean operations, such as negation and conjunction, and functions of natural numbers, such as addition. We can define a standard model for this theory, where sorts and symbols are mapped to their metatheoretic counterparts, with their results wrapped in a singleton set.

When reasoning about this model, we are forced to consider the extended version of application. However, in this case, and with most models defined this way, symbols are either constructors or functional, meaning that they evaluate to a singleton set. Therefore, when these are part of further applications, we can extract this singleton element and defer to non-extended application interpretation.

In the above example, when we reason about a pattern such as $\syapp {\textit{isZero}} {\sephetlist {\syapp S {\sephetlist O *}} *}$, we may utilize the fact that $O$ evaluates to a singleton to simplify the application in $\syapp S {\sephetlist O *}$ to the non-extended version, and since $\syapp S {\sephetlist O *}$ is also a singleton, the application of $\textit{isZero}$ may be simplified as well. In a model with many functional symbols, such as the one in this example, this cascading simplification may be achieved in one step using repeated innermost rewriting.

To facilitate this, we define a predicate over heterogeneous lists of sets that checks whether each element is a singleton set (i.e., for every element, there exists a value such that the singleton set formed with it is equivalent to the element). We then prove a lemma stating that if this condition holds, the extended application is equivalent to the non-extended one with the singleton values extracted. Using this lemma, we can significantly simplify the semantics of pattern applications in most common theories.

We can construct proofs of the aforementioned predicate and apply the simplifying lemma as many times as possible automatically. When combined with automated set reasoning and various other simplification lemmas and procedures, this approach allows us to automatically prove satisfiability for simple models, such as natural numbers or booleans, where
all definitions, including those of sorts and functions such as addition or boolean conjunction, are deferred to their metatheoretic counterparts, and therefore we can leverage the metatheory’s simplification capabilities during the satisfiability proofs.

\section{Conclusion and Future work}\label{sec-conclusion}

In this paper we have outlined a new formalization for the syntax and semantics of matching $\mu$-logic, using dependent type theory as its basis to ensure the correctness of patterns at the time of their construction, as well as to ease the definition of related functions, such as substitutions and extended valuations, that rely heavily on information such as the sort of the pattern and its variable contexts. We have also shown how the newly defined semantics may be used to prove that a model satisfies a theory. We have formalized~\cite{kore-ml} all of this in the \Coq{} proof assistant and highlighted some technical challenges thereof.

In the following paragraphs, we present improvement ideas to incorporate in future developments.

\paragraph{Injection theory}

Besides \emph{definedness}, there is another basic theory that most real-world matching logic specifications rely on: the theory of injections. Since it plays such a crucial role, we plan to integrate it into the syntax of the logic. To accomplish this, we will also extend the signature and introduce a subsorting relation (a partial order) which can be generated from subsorting axioms. This relation will be used in a separate \emph{injection} pattern constructor that can only be used with %
subsorting proofs.
As for the semantics, the model must contain an interpretation of injections, describing how the carrier of the subsort is to be injected into the carrier of the supersort. Furthermore, when defining the carrier of a sort, we extend the metatheoretic type to include constructors for injecting elements of subsorts.

\paragraph{Dependently typed proof system}

It is possible to define a proof system similar to the one described in \Cref{subsec-aml-form}, using the dependently typed syntax. One approach is to reuse that proof system and the existing lemmas by defining conversion functions between dependent and non-dependent patterns, along with additional well-formedness predicates. This would support proving equivalence between the two systems; however, applying this across all existing lemmas would still be cumbersome, and tactics and notations would require significant rework. An alternative is to reimplement the system from scratch, which is cleaner and avoids reliance on prior transformations, but involves substantially more effort.

To ensure soundness in the proof system, patterns must be well-formed. While type indices guarantee most conditions, fixed points additionally require positivity. Encoding this in the type system---by tagging variables with their occurrences---would complicate both construction and type inference, potentially leading to uncomputable types. Instead, we adopt a simpler and already used approach: an external predicate that recursively checks for negative variable occurrences. As this is only relevant when fixed points are involved and not needed for the semantics, it offers a practical balance of correctness and simplicity.

\paragraph{Further automation}

We plan to continue discovering means to automatically generate models and satisfiability proofs for matching logic theories, with a high dependency on the metatheory. Leveraging the fact that carriers are indexed by sorts, we can automatically generate (co)inductive types corresponding to each sort's domain, and carriers defined as mappings to these types. Functions defined via matching logic axioms can similarly be translated into metatheoretic functions using pattern matching. These translations significantly reduce complexity once the axioms are interpreted set-theoretically, enabling many goals to become trivially provable using static proofs and well-designed simplification tactics.

Model and proof generation is intended to be performed by external tooling that outputs \Coq{} code, which is then verified by the proof assistant's compiler. The dependently typed encoding ensures that ill-formed constructs are rejected at compile time, catching errors early whether they stem from incorrect generation or faulty theory definitions. Automated tactics are used to verify model satisfaction; failure indicates either insufficient tactic power, a flawed model, or an inconsistent theory.

We have already created models and satisfaction proofs for some K theories~\cite{kore-ml}, however, this required a lot of manual intervention and the code is still somewhat specialized to these theories. 

\section*{Acknowledgements}

This research was supported by a generous grant received from Pi Squared Inc. Its contents are solely the responsibility of the authors and do not necessarily represent the official views of the entities providing funding for said research.

We would also like to thank Máté Tejfel for his comments and advice in the paper writing process.

\bibliographystyle{eptcs}
\bibliography{biblio}

\clearpage

\appendix

\section{Discussion of Formalization Challenges}\label[appendix]{sec-discussion}

We made a \Coq{} implementation for the syntax and semantics described in this paper. However, we have extended the logic with other common constructs, and faced certain technical difficulties. We discuss the details of the formalization in this section.

\subsection{Extended pattern type}\label[appendix]{subsec-extpat}

While in this article we used the minimal definition of matching $\mu$-logic~\cite{chen2019mu}, in the \Coq{} implementation~\cite{aml-harp}, we have introduced a significant number of connectives as constructors. This makes writing patterns, and simplifying pattern evaluation easier. Note however, that all of these new connectives can be defined as notations on top of the core syntax and the semantics of these constructors is defined to be equivalent to that of their notational equivalents. Many of these new constructs are from first-order logic, as well as a greatest fixed point, and some concepts related to the built-in theory of definedness:

\begin{align*}
	\ofsort{\top}{s} &\coloneqq \exists[s] \ld \ofsortvar{\bevar 0}{s} &
    \ofsort{\bot}{s} &\coloneqq \lnot \ofsort{\top}{s} \\
	\ofsort{\varphi}{s} \lor \ofsort{\varphi'}{s} &\coloneqq \lnot (\lnot \ofsort{\varphi}{s} \land \lnot \ofsort{\varphi'}{s}) &
    \ofsort{\varphi}{s} \to \ofsort{\varphi'}{s} &\coloneqq \lnot \ofsort{\varphi}{s} \lor \ofsort{\varphi'}{s} \\
	\ofsort{\varphi}{s} \leftrightarrow \ofsort{\varphi'}{s} &\coloneqq (\ofsort{\varphi}{s} \to \ofsort{\varphi'}{s}) \land (\ofsort{\varphi'}{s} \to \ofsort{\varphi}{s}) &
	\existentialsort{\forall}{s'} \ld \ofsort{\varphi}{s} &\coloneqq \lnot \exists[s'] \ld \lnot \ofsort{\varphi}{s} \\
	\nu \ld \ofsort{\varphi}{s} &\coloneqq \lnot \mu \ld \lnot \subst{\ofsort{\varphi}{s}}{\ofsortvar{\bsvar 0}{s}}{\lnot \ofsortvar{\bsvar 0}{s}} &
	\total{\ofsort{\varphi}{s}}[s][s'] &\coloneqq \lnot \defined{\lnot \ofsort{\varphi}{s}}[s][s'] \\
	\ofsort{\varphi}{s} =_{s}^{s'} \ofsort{\varphi'}{s} &\coloneqq \total{\ofsort{\varphi}{s} \leftrightarrow \ofsort{\varphi'}{s}}[s][s'] &
	\ofsort{\varphi}{s} \subseteq_{s}^{s'} \ofsort{\varphi'}{s} &\coloneqq \total{\ofsort{\varphi}{s} \to \ofsort{\varphi'}{s}}[s][s'] \\
\end{align*}

\subsection{Computing transports using UIP}\label[appendix]{subsec-uip}

In \Cref{subsec-dependenttt}, we introduced the notion of transports over equalities, and have utilized it several times in the definitions presented in this article. However, in practice these transports can at times become problematic. For example, function applications where the parameter is a value given using a transport over an opaque (black box) proof cannot be simplified.

Let us suppose that we have a function $vlen$, defined by recursive descent on \texttt{Vec}, that tells its length, an opaque proof $H$ that $2 = 1 + 1$, and a vector $xs \coloneqq \app {vcons} 1 {(\app {vcons} 2 {vnil} *)} *$ of type $\app {\texttt{Vec}} {\mathbb N} 2 *$. We may wish to prove that the $vlen$ of $\app {\textit{transport}} {\mathbb N} {(\app {\texttt{Vec}} {\mathbb N} *)} 2 {(1 + 1)} H {xs} *$ (a vector of type $\app {\texttt{Vec}} {\mathbb N} {(1 + 1)} *$) is 2. This makes intuitive sense, as transporting the type of the vector does not change the fact that its length is 2. However, presently, we have no method of progressing with this proof, as $vlen$ is defined by performing case analysis on the structure of its argument vector, which it cannot do, as its argument is a transport.

One axiom regarding equality that can help us compute these specific cases of transports is called the Uniqueness of Identity Proofs (UIP)~\cite{uip}, which states for any two values $x$ and $y$, that if we have two proofs of $x = y$, these proofs must also be equal. If $x$ and $y$ are definitionally equal (i.e. they are the same value after simplifiying all definitions), this also means that the proof must be equal to reflexivity. This is a special case of UIP, often referred to as Axiom K (not to be confused with the $\mathbb{K}$ framework). This gives us the opportunity to replace an opaque proof (such as $H$ above) with a transparent one ($\refl$), which can be used to compute transports, as a transport over $\refl$ is identity.

In the above example, since $1 + 1$ and $2$ are definitionally equal, we may use Axiom K to prove that $H$ is in fact equal to $\refl$, and may be replaced by it in the transport. Since the transport over $\refl$ is the same as the original value, we may remove it, meaning that all that is left to prove is $\app {vlen} {xs} * = 2$, in which we are now able to simplify $vlen$, and see that these are equal by $\refl$.

We note that UIP and Axiom K are sometimes implemented as logical axioms that preserve the consistency of any logic they are added to; however, many dependently typed languages today include these at a lower level, via a concept known as ``pattern matching with K''.

\subsection{Heterogeneous lists}\label[appendix]{subsec-ourheterolist}

In the formalization, we are using a slightly modified version of the heterogeneous list, than we described in \Cref{subsec-dependenttt}. Rather than it being indexed directly by types, we allow it to depend on a list of any type, and also encode a function in the type that maps those values to types.

\begin{typedef}{HList}[\forall (A : \Type) \ld (A \to \Type) \to \listof{A} \to \Type]
    \textit{nil} ! \forall A\ F \ld \app {\typename} A F {[]} *;
    \textit{cons} ! \forall A\ F\ x\ xs \ld \app F x * \to \app {\typename} A F {xs} * \to \app {\typename} A F {(x :: xs)} *
\end{typedef}

The reason for this is that if we mapped the partially applied pattern type to the symbols' parameter sorts in the application constructor, \Coq{}'s type checker cannot ensure that the type is strictly positive. However, if the partially applied type is used directly in the type of heterogeneous lists, it works. The two representations are equivalent, and in practice it does not matter which one we use. In fact, this representation is slightly easier to use in certain cases, as the map function does not need to be simplified.

Regardless of the representation, another issue arises from using the type indirectly, in the argument list of the application constructor. \Coq{} fails to recognize these arguments as being recursive uses of the type when generating induction principles. This makes recursive definitions and inductive proofs over the type difficult, if not impossible.

Fortunately, there is a way to solve this problem. It is possible to define a custom induction principle that includes the induction hypothesis in the application case for every member of the list. Due to the size of the pattern type, this is a lot of code, most of which is boilerplate, however, currently we are not aware of any way to reduce that. This new principle may now be used in place of the old one, where needed.

Another alternative definition we have tried is to use newly introduced variables in the definition of the constructor, and then constrain these to be equal to the list of parameter sorts and return sort of the symbol. However, this solution would not have simplified as well as a computable function, and \Coq{} also failed to recognize strict positivity in the case of one of the equalities, and therefore this approach was also ultimately rejected.

\subsection{Bound variable substitution compared to substitution calculus}\label[appendix]{subsec-nosubtype}

A notable difference between our approach described in \Cref{subsec-subs}, and the substitution calculus described in \Cref{subsec-subcalc}, is that we are not defining it as a set of equalities or a proposition, but rather it needs to be a computable function. This is so that we may use it to define the semantics, as we did in \Cref{subsec-valuations}, which in turn needs to be computable so that we may use it to prove satisfiability. Note that while quotient inductive types may have allowed for a suitable definition, they are not well supported by proof assistants currently.

Initially we have attempted to define the type of substitutions, as they appear in substitution calculus, and use them to define the function, however, since we needed a recursive function, we could not translate the equality where we both have and are substituting a non-zero index, as the substitution is not structurally smaller.

This problem may have been solved by utilizing well-founded recursion, if it is possible to define a measure and a well-founded relation that decreases with the recursive calls. These however, would have needed to take both the pattern and the substitution into account, which introduces a significant amount of complexity, especially since such a pair is already difficult to define for patterns themselves, due to the heterogeneous lists used for application.

Finally, since the parameters for substitution are usually the index and the pattern to be substituted, it would have been necessary to either change the usual interface of the function, or to construct the substitution type from these arguments. Neither of these solutions are ideal.

Because of this, we have decided to investigate if there is a way to define this function directly using the index and the pattern, and while we ran into the same problem of not being decreasing, we have managed to define two new helper functions to substitute just the indices, and to extend the environment of a pattern, that we described in \Cref{subsec-subs}, and with these, it became possible to define the substitution without relying on well-founded induction, and using the original interface.

\subsection{Extended valuation}\label[appendix]{subsec-wfindineval}

In the definition of the extended valuation function mentioned in \Cref{subsec-valuations}, there are a number of cases that are not actually structurally decreasing, and are therefore not accepted by \Coq{}. This is for the existential and least fixed point quantifiers (and by extension, the universal and greatest fixed point quantifiers in the extended syntax), because the recursive call is on a pattern obtained by a substitution on the subpattern.

Because of that, this function is defined using well-founded recursion, over the size of the pattern, with the natural numbers' $<$ relation, which is well-known to be well-founded. The size of a pattern is determined recursively, as the node count of its syntax tree.

In order for this to work, we must prove that $\subst{\varphi}{\bevar 0}{x}$ and $\subst{\varphi}{\bsvar 0}{x}$ are smaller than $\exists \ld \varphi$ and $\mu \ld \varphi$ respectively. Due to the definition of the size of a pattern, this reduces to proving that the substituted patterns have the same size as the original subpatterns.

Because of the way bound variable substitution is defined (\Cref{subsec-subs}), with the help of two intermediate operations, we must reason about these as well, alongside the main substitution function. First, we state that extending the environment of a pattern does not change its size. Next, we prove that substituting a bound variable with a free one yields a size of 1, which is the size a free variable. Finally, with the help of these we can prove the original statement. All of these lemmas may be proven via induction, following the structure of their definitions.

\end{document}